\documentclass[sigconf, nonacm]{acmart}
\usepackage{xcolor}
\usepackage{soul}
\usepackage{xspace}
\usepackage{tikz}
\usepackage{graphicx}
\usepackage{subcaption}
\usetikzlibrary{positioning, shapes.geometric, arrows}
\usetikzlibrary{shapes, arrows.meta}
\usepackage{smartdiagram}
\usesmartdiagramlibrary{additions}
\usepackage{cleveref}
\usepackage{tikzpeople}
\usepackage{algorithm,algpseudocode}
\usepackage[most]{tcolorbox} 
\usepackage{upquote}
\usepackage{csquotes}
\renewcommand{\mkbegdispquote}[2]{\itshape}
\AtBeginEnvironment{quote}{\itshape}

  \everymath=\expandafter{\the\everymath\displaystyle}

\newtcblisting{promptListing}{
  left=1mm,    
  right=1mm,   
  top=0mm,     
  bottom=0mm,  
  colback=gray!10, 
  colframe=blue!50!black, 
  listing only,
  enhanced,
  breakable, 
  boxrule=0.5pt, 
  sharp corners, 
  listing options={
    basicstyle=\ttfamily\footnotesize, 
    language=TeX,
    breaklines=true,
    showstringspaces=false, 
    xleftmargin=0pt,
  },
}
\definecolor{darkblue}{rgb}{0.0, 0.0, 0.5}
\definecolor{darkred}{rgb}{0.5, 0.0, 0.0}
\definecolor{darkgreen}{rgb}{0.0, 0.5, 0.0}
\newtcblisting{pythonListing}{
  left=1mm,    
  right=1mm,   
  top=0mm,     
  bottom=0mm,  
  listing only,
  enhanced,
  breakable, 
  listing options={
    language=Python,
    basicstyle=\ttfamily\footnotesize,
    keywordstyle=\color{darkblue},
    stringstyle=\color{darkred},
    commentstyle=\color{darkgreen},
    breaklines=true,
    breakindent=0pt,
    postbreak=\mbox{\textcolor{darkred}{\footnotesize$\hookrightarrow$}\space},
    showstringspaces=false,
    lineskip=-1pt
  },
  boxrule=0.5pt,
colback=gray!10, 
  colframe=blue!50!black, 
  sharp corners
}

\AtBeginDocument{%
  \providecommand\BibTeX{{%
    \normalfont B\kern-0.5em{\scshape i\kern-0.25em b}\kern-0.8em\TeX}}}

\setcopyright{none}
\copyrightyear{2024}
\acmYear{2024}
\acmDOI{XXXXXXX.XXXXXXX}

\acmConference[Submitted for Review to UIST '24]{UIST}{Oct 13--16,
  2024}{Pittsburgh, PA}
\acmISBN{978-X-XXXX-XXXX-X/24/04}

\begin{document}

\title{Who Validates the Validators? Aligning LLM-Assisted Evaluation of LLM Outputs with Human Preferences} 


\author{Shreya Shankar}
\affiliation{%
  \institution{UC Berkeley}
  \city{Berkeley}
  \state{California}
  \country{USA}}
\email{shreyashankar@berkeley.edu}

\author{J.D. Zamfirescu-Pereira}
\affiliation{%
  \institution{UC Berkeley}
  \city{Berkeley}
  \state{California}
  \country{USA}
}
\email{zamfi@berkeley.edu}

\author{Björn Hartmann}
\affiliation{%
  \institution{UC Berkeley}
  \city{Berkeley}
  \state{California}
  \country{USA}
}
\email{bjoern@eecs.berkeley.edu}

\author{Aditya G. Parameswaran}
\affiliation{%
  \institution{UC Berkeley}
  \city{Berkeley}
  \state{California}
  \country{USA}
}
\email{adityagp@berkeley.edu}

\author{Ian Arawjo}
\affiliation{%
  \institution{Universit\'{e} de Montr\'{e}al}
  \city{Montr\'{e}al}
  \state{Qu\'{e}bec}
  \country{Canada}
}
\email{ian.arawjo@umontreal.ca}


\newcommand{\shreya}[1]{\textcolor{purple}{[Shreya: #1]}}
\newcommand{\ian}[1]{\textcolor{blue}{[Ian: #1]}}
\newcommand{\jd}[1]{\textcolor{orange}{[JD: #1]}}

\newcommand{\evalgen}{\textsc{EvalGen}\xspace}

\newcommand{\topic}[1]{\vspace{-3.5pt}\smallskip \smallskip {\bf #1.}}

\newcommand{\arxivadd}[1]{#1}

\begin{abstract}

Due to the cumbersome nature of human evaluation and limitations of code-based evaluation, Large Language Models (LLMs) are increasingly being used to assist humans in evaluating LLM outputs. Yet LLM-generated evaluators simply inherit all the problems of the LLMs they evaluate, requiring further human validation. We present a mixed-initiative approach to ``validate the validators''---aligning LLM-generated evaluation functions (be it prompts or code) with human requirements. Our interface, \evalgen, provides automated assistance to users in generating evaluation criteria and implementing assertions. While generating candidate implementations (Python functions, LLM grader prompts), \evalgen asks humans to grade a subset of LLM outputs; this feedback is used to select implementations that better align with user grades. A qualitative study finds overall support for \evalgen but underscores the subjectivity and iterative process of alignment. In particular, we identify a phenomenon we dub \emph{criteria drift}: users need criteria to grade outputs, but grading outputs helps users define criteria. What is more, some criteria appears \emph{dependent} on the specific LLM outputs observed (rather than independent criteria that can be defined \emph{a priori}), raising serious questions for approaches that assume the independence of evaluation from observation of model outputs. 
We present our interface and implementation details, a comparison of our algorithm with a baseline approach, and implications for the design of future LLM evaluation assistants. 





\end{abstract}




\received{03 April 2024}

\maketitle

\section{Introduction}
\label{sec:intro}

Large Language Models (LLMs) make mistakes---they hallucinate, ignore instructions, and generate outputs that require  validation~\cite{kalai2023calibrated}. But validating the behavior of LLMs is challenging. In response, researchers and industry developers have created tools for prompt engineering and auditing that help people with testing outputs more systematically~\cite{mishra2023promptaid, jiang2022promptmaker, rastogi2023supporting, arawjo2023chainforge, gero2024supporting, kim2023evallm, promptfoo, kahng2024llm}. 
Such approaches 
require \emph{metrics}---a set of functions to automatically score LLM outputs, each typically an assertion with \emph{true} or \emph{false} values. These metrics increasingly include calls to ``evaluator'' LLMs (e.g., \cite{kim2023evallm, promptfoo, arawjo2023chainforge, zheng2024judging}) that act as ``judges,'' grading outputs on qualities hard to articulate in code; for instance, the ``conciseness'' of an output. 

The problem is that, just like the LLMs they evaluate, LLMs that perform evaluations cannot be trusted. These ``grader'' prompts suffer from the same problems as any other prompt---they are unintuitively sensitive to seemingly minor changes in wording or structure~\cite{sclar2023quantifying}. Yet many existing systems do not include support for verifying the quality of LLM-generated evaluations, asking users to simply trust these outputs. 
How can users reap the efficiency benefits of LLM-assisted evaluation of LLM outputs, while minimizing or avoiding misalignment? How can we help users validate the validators?


In this paper, we propose a mixed-initiative approach, \evalgen{}, to address this automated-evaluation alignment problem in the context of prompt engineering. Our approach streamlines the selection of metrics under practical constraints of user effort and latency. 
Specifically, an LLM suggests criteria in natural language, based on user context (e.g., the prompt under test), that the user can modify. An LLM then generates a pool of candidate assertions for each criterion---either code or LLM grader prompts that output ``true'' or ``false.'' While the user waits for the LLM to generate candidates, they are asked to grade outputs with a simple ``good'' (thumbs-up) or ``bad'' (thumbs-down) voting scheme. \arxivadd{These grades then guide} the automatic selection of assertions \arxivadd{that} optimize for alignment with user preferences (\Cref{sec:algo}). After assertion selection, a final report card reveals \arxivadd{the alignment between the} chosen assertions \arxivadd{and} the user's grades. 
Our approach generalizes beyond the particulars of our specific design, and could be extended to, for instance, update metric implementations with feedback from human preferences, or query the user for finer-grained individual grades. 

\evalgen{} is embedded inside an existing open-source interface for prompt engineering and auditing, ChainForge~\cite{arawjo2023chainforge}. Our alignment algorithm adapts SPADE~\cite{shankar2024spade}, a fully-automated algorithm for generating Python assertions from the revision history of a prompt. We performed an off-line verification of our human-guided alignment algorithm with SPADE as a baseline~\cite{shankar2024spade}, then ran a qualitative user study with nine (9) industry practitioners that use LLMs in production contexts. Because our participants were industry practitioners and thus possibly dealing with NDA-protected data, we offered a task adapted from a real LLM pipeline prompt. Our study design did not impose restrictions on how participants used \evalgen, and users could choose whether to ask the tool to suggest criteria, enter criteria manually, or grade a few LLM outputs first before proceeding onto the criteria specification screen.

Our findings find overall support for \evalgen{}, with one important caveat. We observed a ``catch-22'' situation: to grade outputs, people need to externalize and define their evaluation criteria; however, the process of grading outputs helps them to define that very criteria. We dub this phenomenon \emph{criteria drift}, and it implies that \emph{it is impossible to completely determine evaluation criteria  prior to human judging of LLM outputs}. Even when participants graded first, we observed that they still refined their criteria upon further grading, even going back to change previous grades. Thus, our findings suggest that users need evaluation assistants to support rapid iteration over criteria and implementations \emph{simultaneously}. Since criteria are \emph{dependent} upon LLM outputs (and not independent from them), this raises questions about how to contend with criteria drift in the context of other ``drifts''---e.g., model drift~\cite{chen2023chatgpt}, prompt edits, or upstream changes in a chain. 
Our findings also \textit{(i)} underscore the necessity of \emph{mixed-initiative} approaches to the alignment of LLM-assisted evaluations that also embrace messiness and iteration, and \textit{(ii)} raise broader questions about what ``alignment with user preferences'' means for evaluation assistants. 

We first position our work (Section~\ref{sec:related-work}) and present \evalgen{}'s design (Sec.~\ref{sec:design}) and implementation details (Sec.~\ref{sec:implementation}). We then present two evaluations: an off-line evaluation of our approach (Sec.~\ref{sec:controlled-study}), and a qualitative study with developers (Sec.~\ref{sec:user-study} \& \ref{sec:findings}). Finally, we suggest implications for future work (Sec.~\ref{sec:discussion}).

\begin{figure*}
    \centering
    \begin{subfigure}[b]{0.55\linewidth}
        \includegraphics[width=\linewidth]{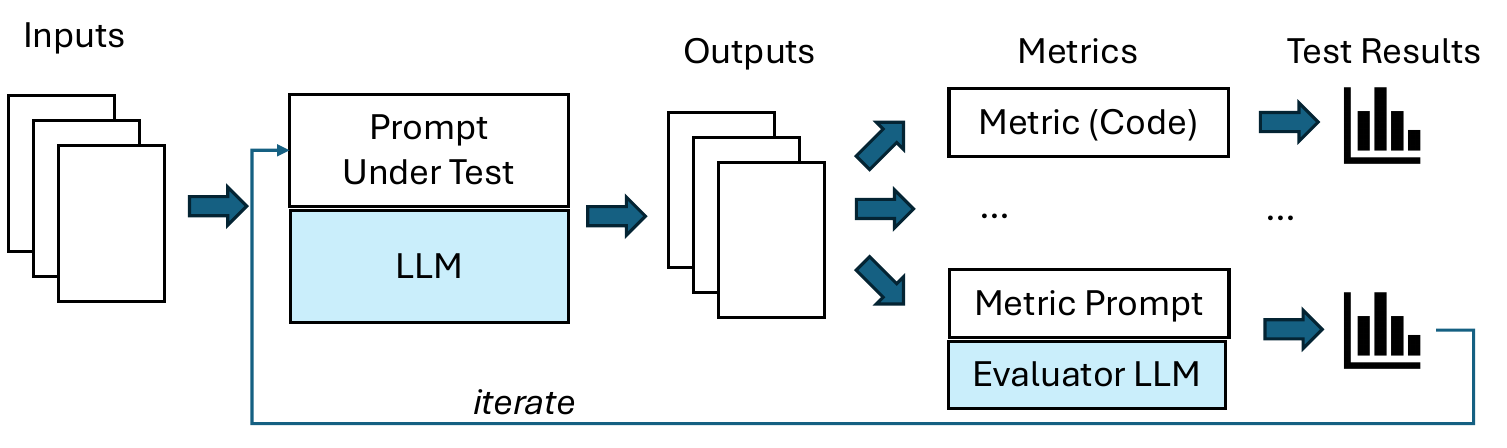}
         \caption{\large Typical Evaluation Pipeline}
        \label{fig:architecture-generalpipeline}
    \end{subfigure}
    \hfill
    \begin{subfigure}[b]{0.65\linewidth}
        \includegraphics[width=\linewidth]{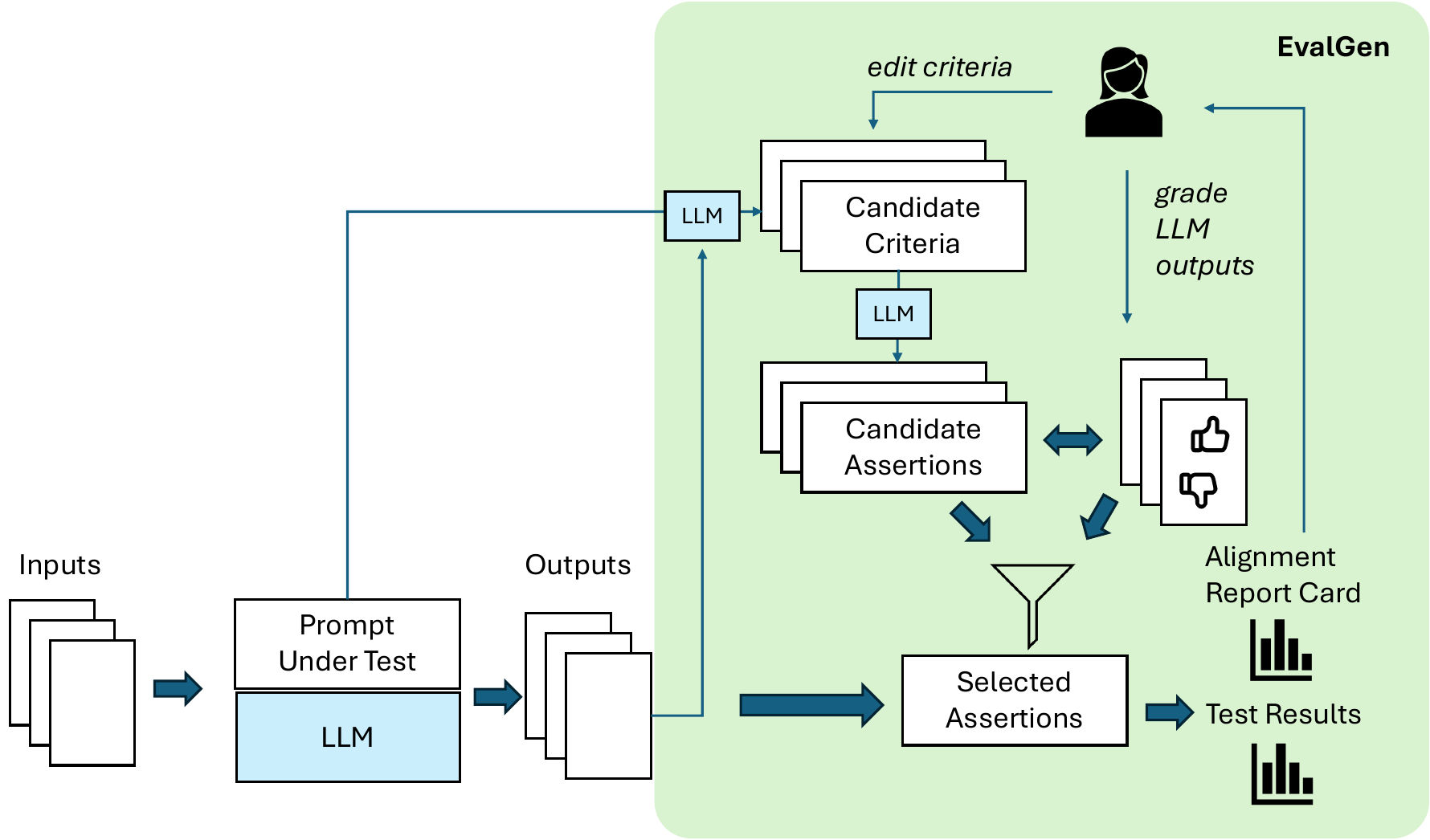}
        \caption{\large The \evalgen Evaluation Pipeline}
        \label{fig:architecture-evalgenpipeline}
    \end{subfigure}
    \Description[Evaluation pipeline diagrams for LLMs]{
Figure 1 depicts two evaluation pipeline diagrams for large language models (LLMs).
(a) The Typical Evaluation Pipeline shows a linear process starting with ``Inputs''
going into a ``Prompt Under Test'' into an LLM, followed by iterations of ``Metrics''
which are coded or prompted, processed by an ``Evaluator LLM,'' and resulting in
``Test Results.''
(b) ``Inputs’' feed into an LLM under a ``Prompt Under Test,‘’ which produces ``Outputs.‘’ The EvalGen Evaluation Pipeline illustrates an iterative process:  the outputs are then processed by EvalGen, which refines ``Candidate Criteria’' and ``Candidate Assertions’' based on how users provide grades on the LLM outputs. The ``Selected Assertions’’ are derived from the grades and candidate assertions, leading to an ``Alignment Report Card’' and ``Test Results.‘’
}
        \caption{\evalgen's approach to assisting users in aligning evaluations. Users iterate through the process of refining criteria and grading. Note that LLM pipeline inputs and outputs are provided by our larger system, and outside the scope of this paper.}
    \label{fig:architecture}
\end{figure*}
\section{Motivation and Related Work}\label{sec:related-work}

In response to the popularity of black-boxed LLMs like ChatGPT, prompt engineering (PE) has emerged as a new practice and research area. Alongside PE is the auditing of model behavior in practices such as ``red-teaming,'' used to identify harmful outputs in internal teams to tweak LLM behavior, usually prior to release \cite[p.17]{liao2023ai}. These tasks have spurred the advent of new tools for ``LLM operations'' (hereafter called LLMOps) and new terminology such as ``prompt template,'' ``chain of thought'', ``chains,'' etc. 

\topic{Automating Evaluations of Prompts} When evaluating LLM behavior, users typically send off hundreds or thousands of queries to models. As users reach the limits of manual evaluation, users set up automated evaluation pipelines (Figure~\ref{fig:architecture-generalpipeline}) in code or with other LLMs (here we use the term LLM-based evaluators; other work uses terms such as ``LLM-as-a-judge'' \cite{zheng2024judging} or ``co-audit'' \cite{gordon2023coaudit}).\footnote{An analog of this problem exists in software engineering as well: developing a set of assertions, often in the form of a set of unit tests or regression tests, that give developers confidence that their code is correct and that code changes do not (re)introduce bugs.} Public PE tools like promptfoo \cite{promptfoo} and ChainForge \cite{arawjo2023chainforge} allow users to write their own evaluation metrics to score LLM response quality, and support both code-based and LLM-based evaluators. For instance, in promptfoo users can write a rubric in a config file to specify how an LLM should evaluate responses, and may use pre-created grader prompt templates or customize them; 
an example is the assertion ``The response is not apologetic.'' Prototypes such as EvalLM~\cite{kim2023evallm} and PromptsRoyale \cite{promptsroyale} also support LLM evaluators, oftentimes exclusively, to help users compare between two prompts. Of PE tools, only EvalLM offers a way to help users calculate the alignment of LLM evaluators with their expectations, but this feature is mentioned only in authors' Design section and is absent from their user study.\footnote{As of this writing, EvalLM is not publicly released.} At best, users of PE tools inspect LLM-generated evaluator outputs manually to double-check; at worst, the tool hides individual scores entirely. \arxivadd{Regardless of aligning implementations of metrics with user preferences, even identifying {\em what} metrics to evaluate for custom tasks remains challenging for LLM practitioners, as evidenced by a recent study~\cite{parnin2023building}. While many evaluation tools require users to declare metrics they care about, some prior work~\cite{shankar2024spade} and \evalgen employ LLMs to propose custom metrics based on prompts in the user's LLM pipelines.}


\topic{Over-trust and Over-generalization of LLM Behavior} That tools provide little assistance to validate evaluator quality is alarming, considering that other research shows people tend to over-rely and over-trust AI systems \cite{liao2022designing,vasconcelos2023explanations,buccinca2021trust,kloft2023ai}. For instance, in one high-profile incident, researchers from MIT posted a pre-print on arXiv claiming that GPT-4 could ace the MIT EECS exam. Within hours, work by Chowdhuri et al. debunked the study \cite{chowdhuri2023no}, citing problems arising from over-reliance on GPT-4 to grade itself. Other work has found further reasons to be cautious: LLMs asked to choose the best response from a set can be consistently biased by set ordering~\cite{li2023split, wang2023large}; 
and LLMs can be highly sensitive to seemingly innocuous formatting changes \cite{sclar2023quantifying}.

A related problem to over-reliance is over-generalization. Zamfirescu et al. \cite{zamfirescu2023whyjohnny} found that users unfamiliar with PE tend to over-generalize from single failures (causing them to throw out potentially good prompts), rather than having a holistic view of the overall performance of a prompt or chain. This was despite the fact that the interface had support for systematic testing. 
Similarly, Arawjo et al. \cite{arawjo2023chainforge} found that even people familiar with LLMs (developers, academics in ML) struggled to scale up their evaluations, appearing to over-generalize from a limited number of outputs even after an automated evaluation pipeline was setup. The authors identified three modes of PE on open-domain tasks, with the second, ``limited evaluation,''' characterized as users ``prototyping an evaluation'' \cite{arawjo2023chainforge}, and suggested that future work focus on supported users in prototyping evaluation pipelines. Over-generalization is common in traditional ML, too---\citet{imperfectai} found that AI systems that showcase subsets of errors, like false positives or false negatives, that have the same accuracy, can lead to vastly different perceptions of accuracy.

\topic{Approaches to Aligning LLMs} The HCI community has extensively studied interactive machine learning (iML). In iML, users iteratively develop models by selecting training examples, labeling data, and evaluating model performance~\cite{dudley2018review}. Interfaces that facilitate seamless transitions between these activities result in fewer errors and outputs that better match users' expectations~\cite{gestalt, simard2014ice}. Some iML interfaces even use ML to assist users, for example, in scaling up labeling---reducing overall user effort required~\cite{desmond2022ai}. When using iML concepts for developing LLM pipelines, we must acknowledge a key challenge with LLMs: they often work with little to no specific training data~\cite{parnin2023building}. Users may simply prototype with inputs they imagine the LLM would see, hoping the prompt generalizes.

In the ML and NLP communities, researchers have explored many ways to align LLMs---and their evaluations---to specific user tasks. Many approaches rely on custom model training or fine-tuning~\cite{christiano2017deep}, but all strategies heavily rely on humans to identify examples of desirable and undesirable outputs. For instance, \citet{liu2023calibrating} demonstrated using annotated LLM outputs---judged on criteria like consistency and relevance---as ``few-shot examples'' for calibrating LLM-based evaluators. Beyond classical summarization and NLP tasks, in response to the ad-hoc tedium of PE~\cite{zamfirescu2023herding}, academics and developers are building automated prompt optimization tools, maximizing some user-defined metric on a labeled set of examples. For instance, given some metrics and prompts, ~\citet{khattab2023dspy} automatically run variations of inserted few-shot examples and LLM-generated rephrasings to optimize the prompt. Other work urges users to write assertions to guide outputs with a mix of code and natural language suggestions~\cite{singhvi2023dspy, rebedea2023nemo}, but writing these assertions is left up to developers, which is often time-consuming and error-prone. A broader point is that research in LLMOps optimization tends to come from the domains of NLP and ML, where authors generally validate tool performance against benchmark datasets with pre-defined metrics, leaving open the question of how well they perform in the wild on idiosyncratic user tasks (e.g. EvoPrompt, PromptBreeder, AutoCalibrate \cite{guo2023evoprompt, fernando2023promptbreeder, liu2023calibrating}). It thus remains unclear how to support developers in their prototyping of evaluations, with the problem becoming even more pressing as the popularity of prompt optimization increases.


Overall, this work reveals that users need more support for (a) \emph{prototyping} evaluations and (b) \emph{validating} evaluators of LLM outputs. It also reveals that auditing LLM outputs is far from easy, with humans prone to the dual biases of over-generalization and over-reliance. One recent LLM-assisted approach, SPADE \cite{shankar2024spade}, makes headway on these issues, helping developers generate Python assertion functions for LLM outputs from prompt history. Here we leverage a similar algorithmic approach to SPADE, but embed it inside an LLM-assisted user interface for evaluator prototyping, \evalgen, that also assists with criteria generation, measuring alignment with human preferences, and visualizing results. 

\section{\evalgen{} Design} \label{sec:design}

\newcommand{\lb}[1]{{(#1)}}

\begin{figure*}
    \centering
    \includegraphics[width=\textwidth]{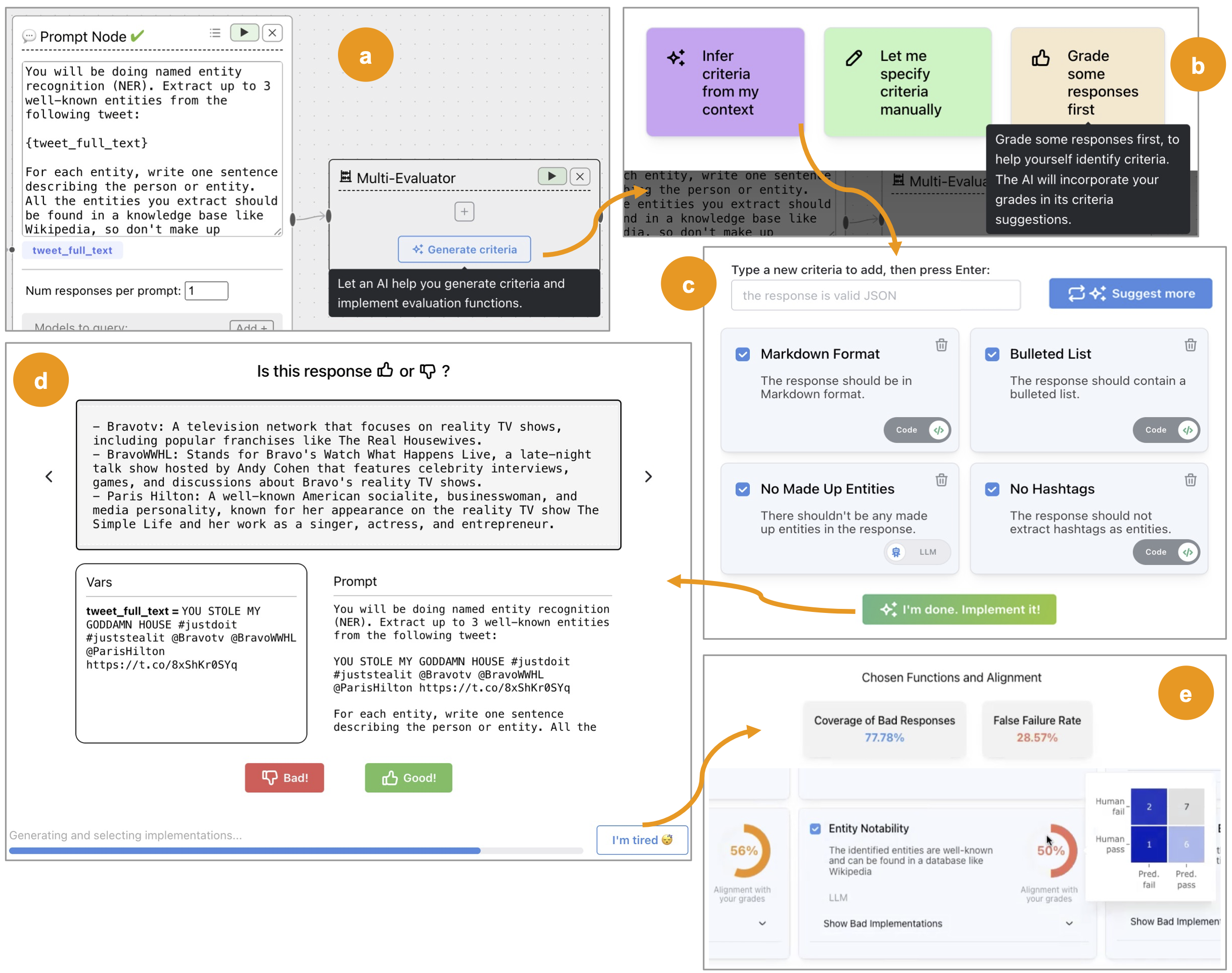}
    \Description[Interactive workflow interface of EvalGen]{
An illustration of EvalGen’s interactive evaluation workflow interface, structured in several panels highlighting different stages of user interaction and feedback. Panel (a) displays the starting point with a Prompt Node and an option to generate evaluation criteria. Panel (b) shows a Wizard interface offering choices to infer criteria, enter them manually, or grade outputs. Panel (c) exhibits a screen for selecting criteria with natural language input and code toggling capabilities. Panel (d) presents a grading interface with options to evaluate outputs as good or bad and an ``I'm Tired'' button to finish the session. Lastly, panel (e) provides a comprehensive Report Card that aggregates the evaluation results and displays alignment metrics through a confusion matrix.
}
    \caption{The workflow of our \evalgen{} prototype, from \lb{a} a Prompt Node attached to an empty Multi-Eval Node, showing a Generate Criteria button; \lb{b} the pop-up EvalGen Wizard with three options, Infer, Manual, and Grade First; \lb{c} the Pick Criteria screen, allowing users to describe criteria in natural language and toggle Code or LLM implementations; \lb{d} the Grade screen, with the LLM output (top), input variables (left), and prompt (right), Good and Bad grade buttons, and an ``I'm Tired'' button (bottom-right) to finish; and finally \lb{e} the Report Card screen, showing the alignment of each criteria and across criteria. Hovering over the alignment shows a confusion matrix. Note that some descriptions and elements have been clipped for space.}
    \label{fig:evalgen-workflow}
\end{figure*}



In designing \evalgen{}, our goal was (1) to investigate how to assist developers in creating evaluators to grade LLM outputs, and (2) to help them ``validate the validators'' through both automated assistance and transparency around how aligned each evaluator is with their expectations. As we covered in Section~\ref{sec:related-work}, emerging practices in prompt engineering, LLM auditing, and prompt optimization involve the writing of evaluation functions (metrics) to automate grading. These functions may be code- or LLM-based. Based on this context, we set out to design an LLM-powered evaluation assistant that provided developers control over metric criteria, evaluator type (code or LLM), and implementation (i.e., function) generation and selection processes, without asking them to come up with criteria or write code or grader prompts themselves.




\subsection{\evalgen{} Workflow}



We implemented \evalgen{} in an existing open-source system for prompt engineering, ChainForge~\cite{arawjo2023chainforge}, which handles querying multiple LLMs with parametrized prompts, running code- and LLM-based evaluators, plotting scores, chaining, etc. \arxivadd{In ChainForge, users write LLM pipelines by creating nodes of various types to represent their dataflow, such as an ``input'' node feeding into a ``prompt'' node.} We discuss here only our extension, chiefly a pop-up screen that helps the user define, implement, and validate evaluation functions. \arxivadd{We also implemented a new node, Multi-Eval, that allows users to include multiple evaluators in a single node and run all evaluators on the outputs of the pipeline's previous node.} Finally, we made improvements to plotting per-criteria scores in the Table View of the LLM output inspector, \arxivadd{which can be accessed via the Multi-Eval node.} 
Fig.~\ref{fig:architecture-evalgenpipeline} provides a high-level overview of the \evalgen{} architecture compared to the typical LLM output evaluation pipeline; we discuss implementation details in Sec.~\ref{sec:implementation}.

{\bf Figure~\ref{fig:evalgen-workflow}} depicts the workflow of in the context of the \evalgen{} interface, excluding returning to the main workflow with selected implementations and using the Table View to inspect scores. \evalgen{} assists a developer in engineering an evaluation of LLM outputs for a single prompt template. First, \evalgen{} is accessed as a button on a ``Multi-Eval'' node we added to ChainForge, which is attached to a Prompt Node ({\bf Fig.~\ref{fig:evalgen-workflow}a}). A Wizard opens, depicting three options ({\bf Fig.~\ref{fig:evalgen-workflow}b}): Infer, Manual, and Grade First. A description of EvalGen (not shown) appears above the options. Clicking Infer or Manual leads to the Pick Criteria screen ({\bf Fig.~\ref{fig:evalgen-workflow}c}); clicking Grade First leads to the Grading screen ({\bf Fig.~\ref{fig:evalgen-workflow}d}) and asks users to grade at least five outputs, before continuing to the Pick Criteria screen. 

The Pick Criteria interface is depicted in {\bf Fig.~\ref{fig:evalgen-workflow}c}. An LLM has generated criteria suggestions in natural language (Sections~\ref{sec:architecture} and \ref{sec:algo}), along with a toggle to prefer a Python code-based or LLM-based evaluator. The user can edit all parts---including the titles or descriptions and type of evaluator---or add new criteria not suggested by the LLM. They can also delete criteria or deselect criteria as needed. Pressing ``Implement It'' passes the criteria to a second LLM that generates candidate implementations. 

While implementations are generated and executed on LLM outputs, users are asked to grade outputs. \evalgen uses these grades to  implementations with their preferences. {\bf Fig.~\ref{fig:evalgen-workflow}d} depicts the Grading screen. A single LLM response is presented to the user, centered in focus in the grader window. The context of the prompt and any input variables (vars) is also present. The user grades outputs with the Good and Bad buttons. 
Since it may be time-consuming to ask the developer to grade on a per-criterion basis, for the grader interface we decided on the simplicity of thumbs-up/down scoring. Such scoring is a noisy yet informative signal of quality---if a response is given a thumbs-up, it is assumed to pass all criteria, and so if a candidate assertion fails on that response, the candidate is down-ranked in the pool (details in Section~\ref{sec:architecture}). Users may also click arrows to navigate through outputs (for instance, if they are unsure about a grade, or want to revise a prior grade).\footnote{Initially, we had asked users to grade a preset number of outputs and a bar showed their progress. However, calling LLMs and executing assertions are asynchronous operations that take an indeterminate amount of time: suggesting an ``end point'' to user grading may lose valuable information when the user still has to wait for generations to return. The user may also find grading enjoyable or important. For these reasons, we did not seek to limit user grading. However, we kept this number-left progress bar in the Grade First screen (accessed via {\bf Fig.\ref{fig:evalgen-workflow}b}).}

Finally, after the user is done grading and all candidate implementations are generated, executed, and filtered for alignment with grades, a Report Card screen appears with feedback on per-criteria and aggregate measures of alignment with user grades ({\bf Fig.~\ref{fig:evalgen-workflow}e}). Hovering over per-criteria metrics shows a confusion matrix of how aligned that particular criterion is to the human grades, while the aggregate metrics show the coverage and false failure rate (see Section~\ref{sec:controlled-study}) of the selected subset of \evalgen{}-generated assertions. The user then returns to the main ChainForge interface (not shown), where the selected implementations are available in a ``Multi-Eval'' node, titled by criteria. The user can edit or add more criteria, inspect and visualize evaluation results (Fig.~\ref{fig:table-view}), etc.; however, this is outside the scope of our design discussion.

Our design reflects trade-offs between developer effort and robust human verification of LLM-generated metrics. The human cannot \emph{completely} validate an LLM-based evaluator: the point of LLM evaluators is to reduce the effort required by the developer, who would otherwise have to grade outputs manually. The only way to fully align an LLM evaluator would be to ask the user to label all outputs; obviously, this defeats the purpose. Asking the developer to grade some outputs, using some time they would have spent waiting anyway, is the key idea behind our design.
\section{Implementation}\label{sec:implementation}

\subsection{System Architecture} \label{sec:architecture}

Like prior work on evaluator assistants~\cite{shankar2024spade, kim2023evallm}, our solution decomposes evaluations into {\em criteria} and {\em assertions} (boolean functions that implement the criteria, evaluating outputs). We employ LLMs in generating criteria, based on the prompt~\cite{shankar2024spade}, and in generating various candidate implementations of each criterion~\cite{shankar2024spade, kim2023evallm}. As users grade, we rank candidate assertions that implement each criteria based on their alignment with user grades (see \Cref{sec:implementation-alignment-def} for how we define alignment). At a high level, alignment is a combination of the assertion's coverage, or ability to fail outputs that the user thinks are bad, and the false failure rate, or ability to not fail outputs that the user thinks are good. \arxivadd{We give a formal definition of assertion alignment in \Cref{sec:implementation-alignment-def}.}


\evalgen's architecture differs from prior work in two main components: first, \evalgen solicits grades from the user on a sample of LLM outputs---requiring some policy to sample LLM outputs to grade. Second, in contrast to SPADE~\cite{shankar2024spade}, which operates offline and solves an integer linear program to generate the optimal assertion set, \evalgen employs an online (i.e., streaming) system architecture to progressively optimize for the most aligned assertion set. When \evalgen generates a new candidate implementation, it immediately executes this implementation on the set of LLM outputs, culling implementations that are obviously bad (e.g., Python functions with runtime errors). \evalgen maintains a dynamic estimate of selectivity (i.e., pass rate) for each candidate assertion, which in turn informs how grades are sampled in the interface. Our system, as depicted in \Cref{fig:architecture-evalgenpipeline}, is structured into three components:

\topic{Criteria Suggestion} We use an GPT-4 to propose various binary evaluation criteria in natural language, such as response length or tone. Developers can select from these suggestions or add their own criteria. For each criterion, developers can select whether it should be evaluated with a purely code-based function or a function that involves calls to another LLM.

\topic{Candidate Assertion Synthesis and Execution} Based on the selected criteria, we use GPT-4 to asynchronously generate one or more candidate assertions as code or a grader prompt to be evaluated by an LLM. \arxivadd{For each criterion, we issue one call to GPT-4 to generate multiple candidate assertions within markers in a streaming fashion. Every time we detect the end of marker in any GPT-4 response, we parse the candidate assertion and submit it to \evalgen's executor, which will run it on LLM pipeline outputs. Generating multiple candidate assertions improves the probability that there is at least one implementation that aligns with developer expectations. Moreover, for code-based assertions, LLMs occasionally synthesize erroneous functions (e.g., hallucinating a function in a Python library), requiring several candidate assertions.}

\topic{Grading Sampler} This component samples LLM pipeline outputs for the user to give binary feedback on (thumbs up/down). When the user grades an LLM output, we update internal estimates of alignment for each candidate assertion, and we sample the next output for the user to grade.

Once the user does not want to grade LLM outputs anymore, or is finished grading all outputs, for each criterion, we select the candidate assertion with the highest alignment with the user's grades. The user can provide a threshold for the false failure rate (as defined in \Cref{sec:controlled-study}) such that \evalgen only selects assertions that do not exceed this threshold.

\subsection{Selecting Assertions \& Eliciting Grades} \label{sec:algo}

Once the user selects criteria, \evalgen's executor begins generating candidate assertions and executing them on all LLM outputs. \evalgen maintains dynamic estimates for the following:

\topic{Selectivity of Candidate Assertions} The selectivity is the probability that an assertion will classify an LLM output as passing. This probability is adjusted each time \evalgen processes the outcome of executing a candidate assertion on an LLM output.

\topic{Confidence Scores for Potentially Poor Outputs} These scores estimate the likelihood that an LLM output is of low quality, without having been explicitly evaluated by the user. The scores are dependent on assertion selectivity and are revised whenever \evalgen evaluates a new assertion against an LLM output, or when a user grades an LLM output directly.

\topic{Assertion Alignment} Alignment, or the harmonic mean of coverage and false failure rate (\Cref{sec:implementation-alignment-def}), is recalculated for each assertion every time the user grades an LLM pipeline output.

See~\Cref{app:algo} for a complete description of assertion selectivity and how it impacts confidence scores; how \evalgen uses these confidence scores to sample grades from the user; formal definitions of alignment; and how \evalgen determines the resulting assertion set based on alignment with grades.

\section{Algorithm Evaluation}
\label{sec:controlled-study}



Here we present results on the effectiveness of \evalgen's selection algorithm. 
Our experiment aimed to understand how {\em soliciting human input} at the criteria suggestion stage impacts the size (number of assertions) and alignment of the resulting assertion set. We compared to a baseline, SPADE~\cite{shankar2024spade}, a fully automated system that generates criteria and candidate assertions and chooses the minimal assertion set that meet coverage and false failure rate constraints. 

\subsection{Evaluation Setup}\label{eval-setup}

We developed two LLM pipelines based on real-world datasets. The \textit{medical} pipeline operates on a dataset of 84  unstructured text transcripts from doctor-patient calls~\cite{yim2023aci}, aiming to extract specific information (e.g., symptoms, medication) without revealing any personally identifiable information (PII). This task requires assertions to ensure compliance with privacy laws. The \textit{product} pipeline involved crafting SEO-friendly descriptions for 100 Amazon products and their reviews~\cite{hou2024bridging}. We selected this task because it mirrors actual LLM applications (there are a number of startups using AI to write SEO-optimized product descriptions), and it benefits from assertions: for example, even if there are negative reviews, the descriptions should not say negative things about the products, which would adversely affect the products' sales potential. Our prompts are presented in Appendix~\ref{appendix:prompts}. For both prompts, the placeholder variables (i.e., \texttt{transcript} and \texttt{document}) represent the input context to inject at pipeline runtime. For each dataset, we executed the corresponding pipeline once for each input using OpenAI's GPT-3.5-Turbo to generate outputs---resulting in 84 medical LLM outputs and 100 product LLM outputs.

Two of the paper authors manually graded all LLM outputs to establish ground-truth labels. Overall, 68\% and 51\% of the outputs were good for the medical and product LLM pipelines, respectively. Common issues included the presence of personal information in the medical pipeline outputs and bad reviews or lengthy content in the product pipeline outputs.

\subsection{Impact of Human Input in the Criteria Generation Step}
\label{sec:controlled-study-spadevsevalgen}

Here, we report quantitative and qualitative differences in SPADE's and \evalgen's assertion sets. There are two differences between SPADE and \evalgen in how they generate assertion sets. The first difference is that \evalgen asks the user to add, edit, or remove criteria before generating different candidate assertions, whereas SPADE does not solicit any input from the user about the criteria. The second difference is in the selection of the assertions themselves: given user-confirmed criteria and a sample of grades provided in a UI, \evalgen picks the most aligned assertion per criterion that meets some false failure rate threshold. Meanwhile, SPADE solves an optimization problem to select a minimal assertion set that meet a false failure rate threshold and cover all SPADE-generated criteria.

\subsubsection{Evaluation Procedure} We first ran SPADE end-to-end for both LLM pipelines, supplying all labeled LLM outputs to see the resulting assertion sets. We initially set both false failure rate thresholds to 10\%. While SPADE met this threshold for the medical pipeline, the product pipeline required adjusting the false failure rate to 40\% to find a viable assertion set. This illustrates the challenge of balancing coverage with false failures, underscoring the need for evaluator systems to effectively navigate these trade-offs.

Subsequently, we ran \evalgen for both pipelines with the same false failure rate thresholds. For the medical pipeline, we defined three evaluation criteria: word count, presence of the six targeted keys, and absence of PII, with the first two implemented via code-based assertions and the last via an LLM evaluator. The product pipeline criteria included absence of negative reviews, absence of links, adherence to markdown format, and word count limitation, with only the first criterion requiring LLM implementation. To create the aligned assertion sets, we provided \evalgen with 16 graded outputs per pipeline instead of all 80-100 graded outputs---given the impracticality of expecting users to extensively grade in a single session.

\subsubsection{Results} Our results in \Cref{tab:spade_evalgen_comparison} show that \evalgen, by incorporating human judgment during criteria selection, achieved equal or better alignment than SPADE with fewer assertions for both pipelines. Specifically, in the product pipeline, \evalgen produced an assertion set less than half the size of SPADE's and increased coverage from 49\% to 73\%. This underscores the benefit of involving humans in selecting criteria, as a fully automated tool may generate assertions for criteria that humans may not actually care about writing assertions for. In the medical pipeline, SPADE added unnecessary assertions, such as one for a neutral tone, not chosen for \evalgen, and split checks for specific keys into more criteria than needed. For the product pipeline, SPADE generated twice the number of assertions compared to \evalgen, some of which were unrealistic, like a Python function designed to flag specific negative phrases such as ``never order'' and ``disappointed'' in the output. In contrast, \evalgen returned a more pragmatic assertion for this criterion---an LLM-based validator to ensure the product descriptions remained entirely positive.

 \begin{table}[ht]
\centering
\begin{tabular}{@{}lcccc@{}}
\toprule
& \multicolumn{2}{c}{\textbf{Medical Pipeline}} & \multicolumn{2}{c}{\textbf{Product Pipeline}} \\
\cmidrule(lr){2-3} \cmidrule(lr){4-5}
\textbf{Metric} & \textbf{\evalgen} & \textbf{SPADE} & \textbf{\evalgen} & \textbf{SPADE} \\ \midrule
Dataset Size & 84 & 84 & 100 & 100 \\
\# Bad Outputs & 27 & 27 & 49 & 49 \\
\# Assertions & {\bf 3} & 5 & {\bf 4} & 9 \\
Coverage & 0.33 & 0.33 & {\bf 0.73} & 0.49 \\
FFR & 0.10 & 0.10 & 0.39 & 0.39 \\
Alignment (\%) & 48.29 & 48.29 & {\bf 66.46} & 54.35 \\ \bottomrule
\end{tabular}
\Description[Comparative table of EvalGen and SPADE]{
Table 1: A comparison of EvalGen and SPADE across Medical and Product pipelines. The table is organized with metrics listed on the left column and the corresponding values for EvalGen and SPADE in the respective pipeline columns. Metrics include Dataset Size, Number of Bad Outputs, Number of Assertions, Coverage, False Response Rate (FRR), and Alignment percentage. EvalGen and SPADE show identical dataset sizes and number of bad outputs across both pipelines. However, EvalGen has fewer assertions, equal or higher Coverage, equal FRR, and higher Alignment in both pipelines compared to SPADE.
}
\caption{Comparison of \evalgen and SPADE Across Pipelines. With user input at the criteria stage, \evalgen achieves the same or greater alignment with fewer functions.}
\label{tab:spade_evalgen_comparison}
\vspace{-20pt}
\end{table}
\section{User Study}
\label{sec:user-study}

To understand how developers might use \evalgen{} to build evaluators for LLM pipelines, we asked nine industry practitioners who had prior experience with LLMs to use \evalgen and think aloud. 

\topic{Recruitment and Participants} We recruited nine industry practitioners via a Twitter post, calling for anyone interested in solving the problem of ``who validates the validators.'' 
We selected the first nine who had experience coding and building LLM pipelines for some company or product. Participants included software engineers, ML scientists, startup executives, and independent consultants. 
We focused on developers with experience with LLMs because they can best intuit what features of an assertion assistant they would find useful, compared to their existing workflows.

\topic{Procedure} All studies were conducted over Zoom. We first spent 5 minutes establishing rapport and asking participants about their backgrounds and experience. We then introduced the participants to our LLM pipeline in ChainForge, where the pipeline's task was to do named entity recognition (NER) on a dataset of tweets. The LLM used in our pipeline was GPT-3.5-Turbo from OpenAI. The prompt for our LLM pipeline was as follows: {\em You will be doing named entity recognition (NER). Extract up to 3 well-known entities from the following tweet: \{tweet\_full\_text\} For each entity, write one sentence describing the person or entity. All the entities you extract should be found in a knowledge base like Wikipedia, so don't make up entities. Return your answer as a bulleted Markdown list, where each bullet is formatted as `- entity: description`. Do not extract hashtags as entities.} We chose this task and set of inputs for three reasons: first, NER is a common, real-world task that language models excel at; second, tweets are short and can be displayed in a UI without any scrolling; third, ``NER for tweets'' is a problem that many people have studied~\cite{gerguis2016asu, liu2013named, suman2021pay}. We allowed participants to change the task or prompt if they wanted. 

Next, we spent a few minutes describing \evalgen's functionality. We showed participants how to view LLM outputs, trigger \evalgen, write an assertion from scratch and add it to their set, run the assertion set on all the LLM outputs, and inspect the Table View plotting results of each assertion on each LLM output. We then gave the participant remote control access to our screen, instructing them to come up with an assertion set for the pipeline. Participants were allowed up to 40 minutes to explore the tool while thinking aloud. We communicated that we were mainly interested in observing their process of creating assertions, not interacting with other features of ChainForge such as comparing different LLM APIs. If the participant had any questions about the interface, we answered them. 

After the participant had an assertion set they liked, or ran out of time, we asked them open-ended questions. We first asked them to comment on \evalgen's approach to generating assertions, and whether they felt that the assertions aligned with their grades. We then asked follow-up questions to learn more about why they felt a particular way or found some aspect of assertion alignment challenging. We limited the post-interview to 10 minutes. At the end, we asked participants to rate how they felt about the assertions' alignment on a 7-point Likert scale (1 = strongly disagree, 7 = strongly agree). Overall, the study ranged from 45min to 1h15min. Our study was approved by our institutional review board (IRB), and participants generously volunteered their time. 

\topic{Analysis} We asked participants to think aloud while using the tool, while we took notes on their thoughts and any visible emotions (e.g., delight when \evalgen suggested a criterion they struggled to externalize, or frustration when they could not find a good assertion for a criterion). We also recorded the transcripts for each video call. We employed open and axial coding~\cite{flick2013sage} to identify common themes across the transcripts and notes for each participant. Initially, we coded individual sentences of interest for each participant, then grouped these into broader themes on a per-participant basis in a second pass of coding. Finally, we consolidated these themes across all participants, reorganizing them as needed.
\section{User Study Findings} \label{sec:findings}

Overall, we found that:

\begin{itemize}
    \item Participants felt that \evalgen was a great starting point for assertions, and wanted to---and could---exercise control over \evalgen's assistance.
    \item Participants struggled to align assertions with their preferences due to two main challenges in grading: {\em (i)} some criteria are difficult for humans to grade (e.g., under a target word count), and {\em (ii)} as they grade more LLM outputs, we observe a {\em criteria drift} phenomenon, in which criteria 
    change as participants grade more LLM outputs (both definitions of existing criteria, and changes to the overall set of criteria). 
    \item Participants' perceptions of alignment and needs varied based on the evaluator type (i.e., code-based vs. LLM-based). 
\end{itemize}

We unpack these findings below. We first describe the typical participant workflow and highlight where participants wanted to exercise control in the process. Then, we discuss the challenges participants faced in aligning assertions with their preferences.

\subsection{Typical Participant Workflow}

All participants ($n=9$) used the provided task (a prompt template for NER of a dataset of 100 tweets, described in \ref{sec:user-study}). Three (3) participants changed the prompt: of these three, one (1) person changed the task from NER to sentiment analysis. After participants had settled on their own prompt (or decided they wanted to use our prompt), each engaged in roughly the following activities:

\begin{enumerate}
    \item {\bf Eyeballing LLM outputs:} Participants viewed the table of 100 LLM outputs, making sure the outputs seemed reasonable at a glance.
    \item {\bf Starting \evalgen:} Participants clicked the button to start \evalgen. This wizard presents three options (Fig.~\ref{fig:evalgen-workflow}b): auto-generate criteria, write criteria, and grade outputs first (before generating criteria). 6 participants clicked the auto-generate button; 1 participant wrote a criterion themselves and then clicked the auto-generate button. The remaining 2 participants wanted to grade first (P4, P9). 
    \item {\bf Grading outputs:} Participants who graded LLM outputs first graded between 5 and 10 outputs, with 2 to 4 ``thumbs-down'' grades. After grading, both participants clicked the button to auto-generate criteria.
    \item {\bf Refining criteria:} After receiving criteria suggestions from \evalgen, participants removed some suggestions and added 1-2 criteria of their own. They usually left evaluation type (code-based or LLM-based) unchanged from what \evalgen suggested, even if \evalgen suggested a type that did not make sense (e.g., checking word count with an LLM API call, rather than code), which happened rarely.
    \item {\bf Grading more outputs:} Participants graded outputs while \evalgen generated and evaluated different candidate assertions. Some participants graded continuously for up to 10 minutes; others stopped after 10 grades.
    \item {\bf Understanding alignment on {\em graded} outputs:} Participants viewed the ``Report Card'' screen, where they all spent a few minutes inspecting the resulting assertion set. Only one participant 
    did not understand what the overall set's coverage and false failure rate meant (P9), but did 
    once the researcher explained it. Participants chose to view the different assertion implementations \evalgen evaluated, including the ones that did not align with their grades.
    \item {\bf Eyeballing alignment on {\em ungraded} outputs:} Returning to the main screen, participants clicked the Run button of the Multi-Eval node to run all assertions on {\em all} 100 LLM outputs and inspected the table of results (Fig.~\ref{fig:table-view}).
    \item {\bf Iterating on criteria:} 3 out of 9 participants again opened the \evalgen wizard to iterate on their criteria and assertions (back to step 2), or stopped because they were satisfied or tired. More than half of the participants expressed interest in iterating on their assertions if they had more time.
\end{enumerate}

\subsection{\evalgen as a Starting Point}

We found that participants liked \evalgen as a {\em starting point} to generating assertions. They felt that exerting control was necessary---since \evalgen sometimes made mistakes---and they did not mind doing so. P8 said:

\begin{displayquote}
This is how I would want a workflow to assist me in evals---basically I want the AI to do 80\% of it, and there can be escape hatches if the AI fails.
\end{displayquote}

Here, we discuss what participants liked and disliked in different components of \evalgen's UI. 

\subsubsection{LLM-generated criteria alleviates writer's block} Eight (out of 9) participants were pleasantly surprised that suggested criteria reflected the criteria they wanted (all except P3). P4 said, \blockquote{I get writer's block when thinking about what assertions to write, so this is great.} P6 said, \blockquote{I feel that if I gave my prompt [for my own pipeline], this would create really good criteria automatically.} Some participants had negative initial reactions: for example, P7 initially expressed dissatisfaction that 7 criteria were generated, as they had expected less. But after deleting some criteria, they realized that the suggestions covered all the criteria they wanted (i.e., they did not have to add criteria themselves), and stated, \blockquote{this auto-generation is sweet}. P3 found that some \evalgen-suggested criteria did not make sense given that assertions only operate on LLM pipeline input-output pairs, like response latency and response stability over time. P3 did not feel that this was a failure of \evalgen, but rather expressed that LLMs are not perfect and stressed the importance of having control over selected criteria.

\subsubsection{Grading outputs first helps refine initial criteria} Participants who graded before selecting criteria (P4 and P9) expressed that they found the grading process very useful. Both participants were happy that \evalgen prompted them to give feedback on why an LLM output was bad. While P9 selected the grading option initially because it seemed like the \blockquote{option that required the least thinking,} they later realized that it was the correct thing to do:

\begin{displayquote}
Of the 3 options, [selecting criteria] probably wasn't the best to start with because I couldn't have extracted [all the] rules directly from the prompt. [While grading, I found] some rules that aren't included in the prompt.
\end{displayquote}

Other participants who did not grade before selecting criteria regretted this decision: for example, P2 said, \blockquote{you should enforce that we all look at at least 20 examples first.} We discuss criteria refinement more in~\ref{sec:grading-shift}.

\subsubsection{Users were happy to grade while waiting} Overall, all participants but one (P7) found grading while waiting for \evalgen to finish generating and executing candidate assertions to be a good use of their time. The participant who did not agree said that they would find it useful if, on the grading screen, \evalgen showed what it was doing with the grades. However, three participants found it difficult to enumerate all criteria in their head and grade against all criteria, wanting \evalgen to prompt them to grade per each criteria (P6, P7, P8). 
Several participants realized they needed to go back and change their grades in \evalgen and were happy they could do this. 

\subsubsection{Users need more control when reviewing alignment on graded results} In the Report Card screen, participants liked that \evalgen generated multiple assertion implementations for each criterion and picked the most aligned one (i.e., highest coverage within a false failure rate threshold of 20\%). However, for some criteria, \evalgen generated no good assertions, and participants deleted the criteria without complaints. P8 said, \blockquote{I like that it tries, because maybe there will be a good implementation!} This suggests that, compared to an opaque approaches we discussed in Related Work, seeing alignment scores helped participants cull assertions that were unaligned rather than blindly trusting them. For criteria where \evalgen did not find any good assertion implementations, participants also expressed interest in writing assertions. Some participants said that they had not graded enough responses to get their ideal alignment numbers, so they wanted to go back and grade more (P1, P7). Participants also liked seeing the coverage and false failure rate of the set, and how each assertion contributed to these overall statistics. However, in this screen, some participants expressed interest in seeing {\em per-criterion} alignment and asked to provide grades for each criterion (P5, P6, P7, P8). Another place where participants wanted to exercise control but were limited by \evalgen's interface was in the selected assertions: P2 and P5 wanted to change a selected assertion to another candidate assertion. Both P2 and P5 wanted to change code-based assertions: for example, P5 said, \blockquote{this is a weird implementation of entity count} and preferred another assertion with a small modification.\footnote{Note that assertions can be edited in the MultiEval node after returning to the main window, but not in EvalGen's report card screen.}

\subsubsection{Users differ in how they assess alignment for ungraded results} 

\begin{figure}
    \centering
    \fbox{\includegraphics[width=\linewidth]{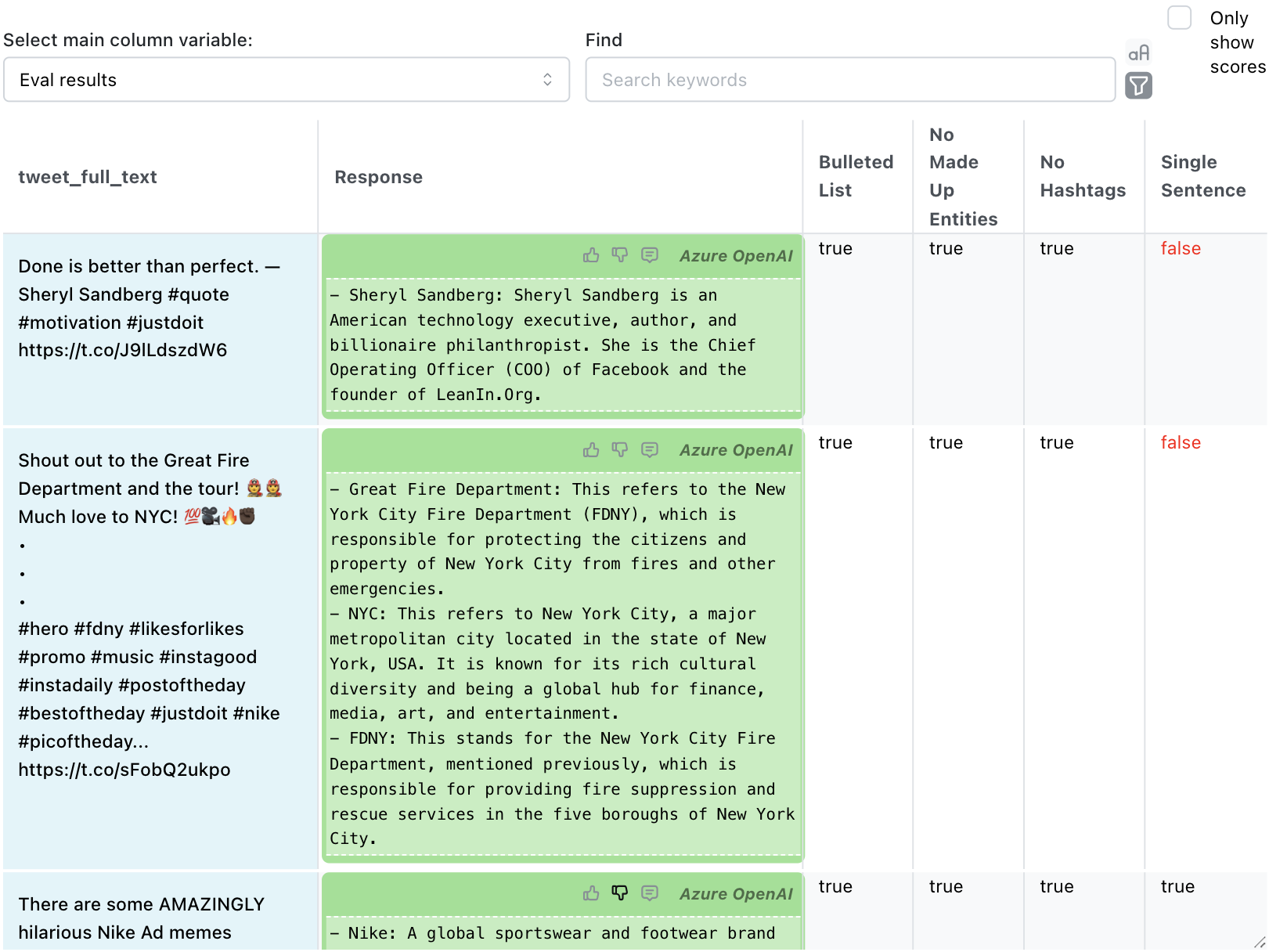}}
    \Description[Table View Interface of EvalGen]{
Figure 3: The Table View, showing the interface for input tweets, Language Model (LLM) outputs, and evaluation results per criteria for the Named Entity Recognition (NER) task. The interface includes a left column with sample tweets and a corresponding right column with LLM responses. The responses are evaluated against criteria such as requiring a bulleted list, disallowing made-up entities, excluding hashtags, and preferring a single sentence structure. Icons indicate whether the response meets each criterion. The figure illustrates an example where the LLM accurately identifies Sheryl Sandberg and the New York City Fire Department but exceeds a single sentence per entity as required by the evaluation criteria.
}
    \caption{The Table View, showing inputs, LLM outputs, and evaluation results per criteria for the NER task (Sec.~\ref{sec:user-study}).}
    \label{fig:table-view}
\end{figure}

Participants {\em really} liked viewing the table of assertion results on all LLM outputs, with all participants expressing interest upon first viewing it~(Figure~\ref{fig:table-view})---but they verified the results differently. In this table, rows represent LLM outputs, and columns represent assertions. P2 said that this table \blockquote{earns trust}; P5 said that it was \blockquote{cool to see all the results on examples [they] didn't grade.} P5 initially did not like using a GUI to come up with assertions, and then said that they \blockquote{prefer looking at this table to [their] current workflow in Jupyter notebooks.} Interestingly, only one participant assessed alignment in this table via coverage---by inspecting each row in the table, one at a time---and everyone else assessed alignment via false failure rate: for each column (assertion), they typically looked for rows (LLM outputs) that the assertion returned False. Some participants wanted this table to automatically recompute, with an accompanying visualization, whenever there were changes to the LLM pipeline; for example, P6 said:

\begin{displayquote}
One thing I would find cool is if there is a way to easily see how changes to my prompt impact the overall [coverage and false failure rate] scores. Just very quickly being able to visualize how [my prompt edit] changes the classifications on a bunch of [LLM outputs.]
\end{displayquote}

This suggests users want data visualization plots between the original prompt and its revision, akin to LLM Comparator and EvalLM \cite{kahng2024llm, kim2023evallm}. Note that ChainForge~\cite{arawjo2023chainforge} also has visualization plots, but we did not introduce participants to this feature.

\subsection{Alignment is Iterative, Criteria- and Implementation-specific}



\begin{table}
\centering
\begin{tabular}{ccccccccc}
\toprule
\textbf{P1} & \textbf{P2} & \textbf{P3} & \textbf{P4} & \textbf{P5} & \textbf{P6} & \textbf{P7} & \textbf{P8} & \textbf{P9} \\
\midrule
6 & 5 & 3 & 4 & 5 & 3 & 1 & 2 & 5 \\
\bottomrule
\end{tabular}
\Description[Participant rating table]{
Table 2: Ratings (1-7, 7 best) for the statement, ``I felt like the assertions aligned with my grades.'' It shows a table displaying the ratings given by nine participants labeled P1 through P9. Participant 1 rated it a 6, P2 a 5, P3 a 3, and so on, with P7 giving the lowest rating of 1.
}
\caption{Ratings (1-7, 7 best) for the statement, ``{\em I felt like the assertions aligned with my grades.}'' Responses were mixed.}
\label{tab:alignment-ratings}
\vspace{-20pt}
\end{table}

Perceptions of the tools' support of alignment were polarized across participants, as shown in \Cref{tab:alignment-ratings}. The main reason why people had low confidence in the assertion report card's alignment with their grades was because they were still uncertain of their own criteria while grading. Participants would say utterances with a questioning tone, like ``I guess'' and ``sure,'' when grading, indicating their uncertainty (P2, P5, P7, P8, P9). Looking more closely at their interactions, we observed a catch-22 situation: participants needed to externalize criteria in order to grade outputs, but they also needed to grade outputs---providing feedback on why bad outputs were bad---in order to externalize criteria. Here, we explore their challenges.


\subsubsection{Criteria drift}\label{sec:grading-shift} Grading outputs spurred changes or refinements in participants' criteria, which we refer to as criteria drift. We observed two types of drift.

First, participants wanted to {\em add new criteria} when they observed new ``types'' of bad LLM outputs (P2, P5, P6, P8, P9). In the \evalgen{} interface, they could not go back and add new criteria; they had to wait for all candidate assertions to finish executing, move past the report card screen, and start a new \evalgen process. 

Second, as participants graded more outputs, we found that they {\em 
reinterpret existing criteria to better fit the LLM's behavior} (P2, P5, P6, P8, P9). For example, P2 and P8 had a ``proper noun'' criterion, which was supposed to assess that ``the entities extracted were proper nouns.'' At first, they rated as bad any LLM outputs that contained {\em any} entity that was not a proper noun. But, after observing that responses had varying numbers of proper nouns, both wanted to change their criteria such that {\em most} of the entities were proper nouns, rather than all. P9 appreciated the ability to provide feedback on unsatisfactory outputs before finalizing the initial criteria set, as this process helped them articulate their criteria. Twice, P8 mentioned that they gave a bad grade not because they believed the output was bad, but because they wanted to be consistent with previous grades---good labeling practice, perhaps, but not good for alignment. P7 noticed that in some outputs included hashtags from the original tweet inputs (e.g., \#justdoit), while in other cases, the outputs did not use the hashtag symbol but still mentioned the entities referred to by the hashtags (e.g., ``Nike'' instead of \#Nike). This inconsistency led P7 to rethink the criteria definition for including hashtags; specifically, whether it was acceptable for the LLM to replicate entities from hashtags, provided the hashtag was removed. P5 expressed the same uncertainty. \blockquote{I think it's hard to know until you see it,} P7 said.  

\subsubsection{Users prefer to adjust their grading approach based on the difficulty of evaluating a criterion} Overall, participants generally liked the process of grading LLM responses and feeling like the grades were useful, but they wanted to prioritize grading criteria they felt {\em needed their alignment}---especially for LLM-based assertions (P3, P5, P7, P8). For example, P3 expressed that they would trust the assertions more if the \evalgen process allowed them to set different false failure rates per criteria (since LLMs might be bad at evaluating some criteria), instead of one global false failure rate constraint for the entire assertion set:

\begin{displayquote}
There are criteria where you can be okay with failing, and then there are other criteria where you are like, `this must absolutely pass'... 
[T]here's a [spectrum] of failure as opposed to: it just passes or fails.
\end{displayquote}

Relatedly, some participants expressed that they didn't trust their grades because they {\em themselves} couldn't evaluate some criteria as well as an automated solution (P2, P5, P6, P7, P8; we discuss further in \Cref{sec:code-vs-llm-evals}). A criterion like word count is hard for humans to assess but easy for a good Python function to evaluate. P8 desired to grade for only one criteria, reasoning that it might improve efficiency (\blockquote{I generally want to be in the loop for these tests...but I want to put myself in the loop in a way that is efficient.}).

\subsubsection{What constitutes ``alignment'' is subjective, especially when converting natural language criteria to code-based assertions}\label{sec:code-based-misinterpretation} \arxivadd{For code-based assertions, \evalgen's interpretation of the criterion (i.e., GPT-4's interpretation) did not match what the participants expected. As such, no matter how many grades the participant gave, all candidate assertions were similarly misaligned. Participants who observed this were confused why their grades seemingly had little impact on some of the chosen assertions (P3, P5, P7, P9). For example, while all participants had a criterion to enforce that there were no entities with hashtags in the output, some participants interpreted this as any hashtags representing entities should not be extracted as entities: e.g., if the output included the hashtag \#Nike, P5 did not want Nike to be extracted at all. On the other hand, P9 wanted Nike to be extracted as an entity, but they did not want the LLM output to include the hashtag. Both P5 and P9 got the same code-based assertion for the criterion, which simply checked for the presence of the hashtag character in the output---this assertion did not align with P5's grades, but did with P9's. This particular misalignment can also be viewed as an instance of criteria drift, as described in \Cref{sec:grading-shift}, since P5 was only able to refine the criterion after grading several LLM outputs. For another criterion, P5 felt that \evalgen could not find a good assertion that aligned with their grades, but also said that they were \blockquote{lost at what would be a good implementation.}} 

\arxivadd{Overall, alignment is not merely a matter of performance (i.e., the idea that ``a better LLM would do better''): we observed that misalignment sometimes occurred due to tacit criteria that participants held which was not explicitly explained in natural language. Like prior work has found for cross-LLM comparison \cite{arawjo2023chainforge}, this tacit understanding of criteria could be highly subjective and contradictory across participants.}

\subsection{Alignment Needs and Preferences Differ for Code vs. LLM Evaluators}\label{sec:code-vs-llm-evals}

All participants liked \evalgen's ability to assist in generating both code-based and LLM-based assertions, expressing the need for both types of assertions. Both P4 and P6 mentioned that they liked the ability to correct \evalgen's suggested type. Yet participants reasoned about code-based assertions and LLM-based assertions differently: while they liked having assertions of both types, they wanted to construct and iterate on them differently in order to feel like the resulting set of assertions aligned with their preferences.

\subsubsection{Users like having control over the evaluation type and thought each evaluation type had a different affordance} Participants seemed to have a clear idea of when a criterion should be evaluated by code, and when it should be evaluated by an LLM. Generally, for formatting checks (e.g., asserting that an output is in Markdown), count-based checks, and checks to include or exclude specific phrases, participants preferred code-based assertions. P6 and P8 expressed that they wanted to use LLM-based assertions for their own LLM pipelines because \blockquote{most of their criteria was fuzzy.} 
P2 selected criteria to be evaluated via LLM whenever they could not immediately think of a Python function that could implement that criteria. We also observed a participant (P8) preferring LLM-based to code-based assertions: LLMs are relatively forgiving when encountering outputs with an unexpected quality or format, whereas code assertions check exact constraints. 

\subsubsection{Users can---and want to---directly verify code-based implementations} Regardless of the type of assertion (code or LLM), \evalgen follows the same process of exploring multiple candidate assertions per criterion and, for each criterion, selecting the assertion that aligns most with the user's grades. While participants generally liked this approach for LLM-based assertions, they did not like this for code-based assertions. This observation may be specific to users who are experts in coding (as all of our participants are). Some participants wanted to see the code for each candidate assertion and select the best Python function for each criterion themselves (P2, P5, P7). P5 said:

\begin{displayquote}
When something can be solved using Python code, I do have an envisioned [implementation] in mind that I can easily verify. Just showing [me] the [code] will be quicker.
\end{displayquote}

P7 wanted to iterate on code-based assertions simply by providing feedback to make the assertion more or less ``fancy''---for example, in checking for a pattern in the LLM output, they might want the regex to be more or less complicated depending on the sample of LLM outputs they've seen. However, P7 then acknowledged that their ability to edit depends on code length and complexity:

\begin{displayquote}
For something that's like 10 lines of code or less, I'll look at [the code]. But for [longer functions,] I probably want to eyeball [the function results]. This is where it gets a bit unclear.
\end{displayquote}

While participants had lots of ideas for improvement, participants overall felt that \evalgen provided a good start to code-based assertions, and it was easy for them to edit the code if they wanted: two participants even asked for the ability to export code-based assertions as unit tests or in a Python file (P4, P6). 

\subsubsection{Users want their grades to aid automated criteria generation} While participants liked that \evalgen tried multiple candidate assertions for each criterion, they wanted \evalgen to use their grades, as well as natural language feedback on thumbs-down grades, in {\em prompting} LLMs for candidate assertions (P3, P4, P5, P7). For example, any time a user gives a grade, \evalgen could query an LLM for a new candidate assertion, including the new grade in the prompt so that the candidate assertion may be more aligned with the user's grades. This strategy can, of course, be used to generate new code-based assertions too. Not only did participants want grades to be included in the process of generating new candidate assertions, but some wanted their labeled LLM outputs to be included in the LLM-based assertions' prompts themselves (P3, P5). P3 excitedly realized this when looking at the table of results (\blockquote{I can take these good and bad examples and use them as few-shot exemplars in the prompt for the LLM evaluators...}).
This suggests an optimization loop akin to ConstitutionMaker and DSPy \cite{petridis2023constitutionmaker, khattab2023dspy}, but for \emph{assertion generation and validation}, rather than prompt optimization.

\subsubsection{LLM-based assertions are harder to trust} While participants found \evalgen's suggested code-based assertions to be more obviously misaligned than the LLM-based assertions (\Cref{sec:code-based-misinterpretation}), they also acknowledged that LLM-based assertions are harder for them to trust---since they can edit the code-based assertions more easily than the LLM-based assertions (P2, P5, P6, P8, P9). 
Some participants were skeptical of how LLM-based assertions might transfer to monitoring LLM outputs in a production pipeline (P3, P6, P8). P8 said, \blockquote{I cannot begin to think about how LLMs as validators [in production] can work, I’m very skeptical.} P9 asked, \blockquote{How do I maintain my evals over time; do I have to rerun this entire process?} One suggestion is an interface that solicits grades from other people---possibly the end-users of the user's LLM pipeline---to continually realign LLM-based assertions.

\section{Discussion}\label{sec:discussion}

Here, we unpack criteria drift and how it affects alignment, discuss the implications of our findings for future LLMOps evaluation assistants, and outline some open questions.

\subsection{Implications of Criteria Drift for LLM Evaluation Assistants}

The practice of benchmarks in ML and NLP presume a world of well-defined criteria (and well-labeled data) on which to judge LLM outputs. For instance, \textsc{AutoCalibrate} is a method to calibrate LLM evaluators with human preferences that requires large expert-labelled datasets with settled (i.e., established upfront) criteria~\cite{liu2023calibrating}. However, in practice, we found that developers rapidly iterate over criteria, and furthermore that cognitively engaging with LLM outputs helps them to refine their criteria. This suggests criteria refinement and grading should happen {\em in tandem} in interactive settings, and poses challenges to alignment methods that presume settled, expert labels. In particular, the dependence of criteria on a \emph{specific prompt and LLM's} outputs means that any change to the LLM pipeline can cause criteria drift: for example, LLM APIs can be replaced with new models, resulting in prompt drift~\cite{chen2023chatgpt}.

Future system designs should support these requirements. For instance, an evaluator assistant might adjust criteria dynamically as the user grades and gives feedback. We might also consider per-criteria grading (while keeping the simplicity of thumbs up/down voting). Eliciting grades at the criteria level might allow evaluation assistants to more precisely adjust both the criteria for evaluation, as well as the way these criteria are implemented. Finally, including examples of both good and bad LLM outputs within the LLM-based evaluator prompts could also be beneficial (a method akin  to recent work~\cite{khattab2023dspy, madaan2024self}). However, whenever outputs change, \arxivadd{we may need to ask users to re-grade or re-think their criteria.} 

Our criteria drift finding echoes prior work in educational settings, where instructors often update their grading rubrics to reflect common errors as they grade more assignments~\cite{singh2017gradescope}. Similar to how teachers adjust rubrics and regrade as they see more student submissions, making users reassess LLM outputs after updating criteria could enhance the accuracy and consistency of grades. Given the observed inconsistency in participants' grading, evaluation assistants might also pursue crowdsourcing methods like majority voting and self or external assessments to determine accurate grades for LLM outputs~\cite{davani2022dealing, dow2012shepherding}. Another challenge for evaluation assistants is extending adapted criteria to grade both {\em ungraded} and future {\em unseen} LLM outputs. How do evaluation assistants consistently sample grades for outputs that reflect the overall distribution of LLM pipeline successes and failures?  

The reader might wonder when criteria ``settle.'' Perhaps there was simply not enough time in our study, and had participants graded for an hour or two, they might have solidified their criteria, and criteria drift goes away.
There is reason to believe that this situation does not change with more time---as we saw, the criteria participants refined changed \emph{to adapt to the behavior of the LLM outputs being evaluated}---a \emph{dependent,} rather than independent assessment of quality. In the real world, similar situations exist where criteria are never ``fully settled'' as more inputs come in---consider the court of law. One of our participants remarked that people ``know a bad output when they see it.'' Their adage reflects a U.S. Supreme Court Justice's famous opinion in a 1964 court case about obscene content~\cite{gewirtz1996know}. As in that remark, the decisions of human validators seem at first glance ``to be based on a non-rational, intuitive gut reaction, instead of reasoned analysis; it seems to be utterly subjective and personal''~\cite[p.1025]{gewirtz1996know}. 
However, perhaps, as the law scholar Gewirtz argues, subjectivity is \emph{not} necessarily a sign of irrationality (contrasting with some imagined future AI that is entirely objective and rational, entirely ``aligned'' or ``better'' than humans). On the contrary: ``There are good reasons to accept the imperfect in a judge. We should encourage judges to believe and say: This is the best I can do now; it doesn't solve all the problems, but it's a start, and I'll keep thinking''~\cite[p.1027]{gewirtz1996know}. This raises a deeper epistemic question for evaluation assistants---is ``alignment'' an actualizable goal? 
To what extent does our common terminology and assumptions---e.g., that there is a ``ground truth'' set of labels we merely need to elicit---fail us? Is validating the validators only ever a work-in-progress? 


\subsection{Operationalizing Assertions}


Participants expressed the desire to deploy their assertions in production. Some wanted to deploy assertions in the critical path of the LLM pipeline, to catch bad outputs before they are shown to end-users. Others wanted a more passive deployment---one participant requested for some form of report sent to their email inbox daily, containing results of the assertion set executed on a sample of LLM outputs from that day. In practice, assertions have different operational requirements. For example, if an assertion runs in the critical path, it should not result in many false failures, otherwise the LLM may get unnecessarily reprompted (i.e., rerun) and the overall latency might increase for our LLM pipeline. Moreover, as criteria changes over time, evaluation assistants should ask users to grade new LLM outputs observed in production and automatically adapt assertion sets.

Our study confirmed the dual necessity for both code-based and LLM-based assertions among participants, but further showed that participants felt each type required distinct treatment, both in implementation selection and in how they perceive alignment. Often useful for sanity checks like output structure, code-based assertions can be used in the critical path of the LLM pipeline. In fact, a number of LLMOps tools exist for users to implement such code-based guardrails~\cite{rebedea2023nemo, guardrails, instructor}. However, our participants highlighted the challenge of finding the {\em right} implementation for a given assertion---a decision that can depend intricately on the characteristics of LLM outputs. For example, determining the acceptable length of a response might vary significantly based on the observed output distribution. In fact, specifically for an output length criterion, P3 mentioned that they would look at a histogram of word lengths for a batch of outputs and see if there were some buckets of anomalous word lengths that exhibited other failure modes (e.g., a ``rambling'' response from the LLM). 

Some participants mentioned that they wanted their \emph{collaborators,} such as product managers, to grade outputs in \evalgen. Importantly, when allowing multiple users to collaborate on grading, evaluation assistants have to consider inter-rater reliability and handle disagreements, if any. Again, we can draw inspiration from techniques from the crowdsourcing literature: for example, evaluation assistants could model each individual grader's accuracy and apply corrections if necessary~\cite{zhuang2015debiasing}. Moreover, collaborators may have different experience levels with coding: those with less experience may choose LLM-based assertions for criteria that might be better evaluated with code. While \evalgen provides initial suggestions for whether criteria should be evaluated with code or another LLM, evaluation assistants may want to make sure that all users from the same team ``agree'' on the right implementations. One could imagine interfaces similar to creating a ``pull request'' for a new assertion and soliciting review from a team member, and a workflow similar to continuous integration/continuous deployment (CI/CD) that seamlessly pushes new assertions to production.

\subsection{Future Work and Limitations}

Assertions serve as a straightforward mechanism for evaluating LLM outputs, yet the potential of evaluation assistants extends beyond \evalgen's supported binary judgments. Consider the example of word count: rather than setting rigid thresholds, one might find it more useful to monitor variations in word count across different LLM outputs. Like in traditional ML monitoring~\cite{shankar2023automatic}, there is, of course, still imprecision in what word count counts as ``bad,'' given its distribution. Does the AI assistant's definition of bad align with the user's? Tracking finer-grained information in evaluations, beyond simple true/false conditions, can aid in debugging issues within LLM pipelines, once a user knows the output is bad. Second, many users write their LLM pipelines as chains of multiple LLM calls~\cite{arawjo2023chainforge, wu2022promptchainer, LangChain, flowise}. Evaluation assistants can facilitate end-to-end alignment, which can be complicated not only with chains of LLM calls, but also in compound AI systems that leverage other components like retrieval-augmented generation~\cite{compound-ai-blog}.

Increasingly, user want their prompts to automatically improve based on assertion results. For instance, if an assertion fails for a group of LLM outputs, one might want to add an instruction to the prompt to solve this failure mode. Some frameworks already experiment with using feedback from assertions or user grades to refine prompts~\cite{singhvi2023dspy, madaan2024self, petridis2023constitutionmaker}. However, incorporating this into an evaluation assistant is not as straightforward: there may be a cyclical process where evaluation assistants not only help in adjusting assertions based on prompt performance but also use these insights to suggest further prompt modifications. This iterative cycle of evaluation and adjustment could foster a co-evolutionary environment where prompts, assertions, and even evaluative mechanisms themselves are continuously refined in a unified interface.

Finally, there are two limitations of our study that are worth pointing out. First, our off-line evaluation only focused on two pipelines. Different applications may warrant different methods of sampling grades and aligning assertions. Second, our qualitative study focused on a small sample of developers, all of whom have significant experience deploying LLMs. Moreover, participants did not have enough time to iterate many times in \evalgen, and our setup did not cover the deployment phase of LLM workflows. Future work might explore best practices and pitfalls of evaluations in the broader LLMOps lifecycle.
\section{Conclusion}

This work presented \evalgen, a mixed-initative approach to aligning LLM-generated evaluation functions with human preferences. \evalgen assists users in developing both criteria acceptable LLM outputs and developing functions to check these standards, ensuring evaluations reflect the users' own grading standards. In a qualitative study with 9 expert users, we observed a pattern we call {\em criteria drift}, where users refine their evaluation standards as they grade more LLM outputs. Recognizing the dependency of criteria on LLM outputs highlights new directions for designing future evaluation assistants.

\bibliographystyle{ACM-Reference-Format}
\bibliography{sample-base}

\clearpage
\appendix

\section{Algorithms for Selecting Assertions \& Eliciting Grades} \label{app:algo}

\subsection{Assertion Selectivity and Impact on LLM Output Quality Confidence}\label{sec:implementation-sigma-def} One way to establish confidence in whether an LLM output is problematic is to assess the selectivity, or pass rate, of assertions that fail it. Intuitively, assertions that frequently fail outputs (low selectivity) provide limited insight into output quality. For example, an assertion that trivially fails every output offers no discernment and has a selectivity of 0. 

\evalgen leverages selectivity estimates of assertions to assign a confidence score to each LLM output, indicating the likelihood it is of poor quality. The rationale is straightforward: an output is more likely to be problematic if failed by assertions known for their high selectivity. Concretely, for a set of assertions $F$ where each assertion $f \in F$ returns 1 for a pass and 0 for a fail, we calculate the confidence score for an LLM output $e$ as follows:

\begin{align*}
\sigma\left(e\right) = \sum_{f \in F} \text{selectivity}\left(f\right) \times f\left(e\right)
\end{align*}

The score $\sigma$ is always non-negative. A score of 0 means no assertions have failed the output, indicating a higher likelihood of quality, while lower scores, resulting from failures by non-selective assertions, point to uncertainty or potential issues with the output.

\subsection{Sampling Grades} \label{sec:sampling-grades} Given that users may not want to grade so many outputs in the \evalgen interface, choosing which outputs for users to grade is crucial for aligning the system's evaluations with user expectations. Randomly selecting outputs without considering their predicted quality can lead to misalignment, especially if the selected samples aren't representative of the entire dataset. Prior work also underscores the importance of soliciting a representative graded sample of LLM outputs~\cite{boyeau2024autoeval, shankar2024spade}.

Given $\sigma$ scores as previously defined, we consider a number of strategies to sample outputs for grading:

\begin{itemize}
    \item {\bf Random:} Sample outputs at random (uniformly)
    \item \textbf{Highest:} Sample the outputs with the highest $\sigma$. This approach focuses on potentially problematic content.
    \item \textbf{Lowest:} Sample the outputs with the lowest $\sigma$, prioritizing outputs that don't fail any assertions or fail low-selectivity assertions.
    \item \textbf{Alternating:} Alternate between high and low $\sigma$, aiming for a diverse sample with both bad and good outputs.
\end{itemize}

In \Cref{sec:controlled-study}, we test these strategies against a random baseline on two different LLM pipelines. We employ an alternating sampling policy for the \evalgen user studies. We do not claim to have the best sampling policy; we chose an alternating policy with the hope that it would solicit a balanced sample of good and bad grades.

One may wonder why we do not list a policy that ranks the outputs by score and samples the middle for grading. While this might seem akin to seeking out uncertain cases---as is common in active learning---our scores represent the likelihood of outputs being poor. They do not differentiate between good and bad per se. Therefore, outputs with low scores may still vary widely in quality, reflecting our system's uncertainty.

\subsection{Choosing Aligned Assertions}\label{sec:implementation-alignment-def} Once the user stops grading and wants to see the alignment Report Card, as depicted in \Cref{fig:evalgen-workflow}e, for each criterion, we select the candidate assertion with the highest alignment score. We adopt notation from \citet{shankar2024spade} in defining alignment. Formally, let $E$ be a set of LLM pipeline input-output pairs and $f: E \rightarrow \{0, 1\}$ represent an assertion. Let $y$ be a binary vector, where $y_i \in \{0, 1\}$ represents whether the user thinks an LLM output $e_i$ is bad. Suppose $F = \{f_1, f_2, \ldots, f_j\}$ is a set of $j$ assertions. Concretely, the coverage and false failure rate (FFR) of $F$ is represented by the following equations:

\begin{align*}
\text{Coverage}\left(F\right) = \frac{\sum_{i} \mathbb{I}\left[y_i = 0 \land  \left(\exists f \in F, f\left(e_i\right) = 0\right) \right]}{\sum_{i}  \mathbb{I}\left[y_i = 0\right]} \\
\text{FFR}\left(F\right) = \frac{\sum_{i} \mathbb{I}\left[y_i = 1\land \left(\exists f \in F, f\left(e_i\right) = 0\right) \right]}{\sum_{i} \mathbb{I}\left[y_i = 1\right]}
\end{align*}

In both definitions, $\mathbb{I}$ is the indicator function. Intuitively, coverage represents the set's true negative rate, while false failure rate represents the set's false negative rate. An aligned set of assertions would have a high coverage and low false failure rate. We define the alignment of $F$ as the harmonic mean of coverage and the inverse of FFR:

\begin{align*}
\text{Alignment}\left(F\right) = 2 \times \frac{\text{Coverage}\left(F\right) \times \left(1 - \text{FFR}\left(F\right)\right)}{\text{Coverage}\left(F\right) + \left(1 - \text{FFR}\left(F\right)\right)}
\end{align*}


Note that alignment is very similar to F1 score; however, we are concerned with the precision and recall of {\em failures} (i.e., when $f = 0$, not when $f = 1$), and we are concerned with a set (i.e., when {\em any} assertion returns 0).

\subsection{Evaluation of Sampling Policy}


We described in~\Cref{sec:sampling-grades} four options we considered to sample LLM outputs: random, highest, lowest, and alternating. Here, we compare \evalgen's sampling policy, \emph{alternating}, to these other three baselines. For this experiment, we sampled 16 outputs to grade, but in practice the user can grade more or fewer.  Using the same LLM pipelines as described in \Cref{eval-setup}, to assess sampling variance, we conducted 10 trials for each of the four sampling policies---where, for each trial, we kept the same set of candidate assertion functions.

\begin{figure}
    \centering
    \resizebox{\linewidth}{!}{\input{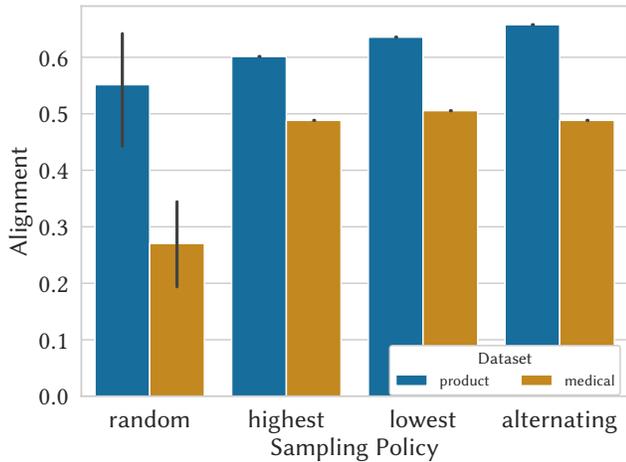}}
    \Description[Bar chart of alignment for different sampling policies]{
Figure 4: A bar chart showing the alignment for assertion sets resulting from different sampling policies to grade LLM outputs. The chart has four groups of bars representing random, highest, lowest, and alternating sampling policies. Each group has two bars, one for the product dataset and one for the medical dataset, indicating the alignment score. The product dataset bars are colored blue and the medical dataset bars are colored yellow. The chart illustrates that random sampling introduces significant variance in alignment, with the ``alternating'' policy generally showing higher alignment, and the ``lowest'' sampling policy showing lower alignment for both datasets.
}
    \caption{Alignments for assertion sets that result from different policies to sample grades from the user. Each policy was tested across 10 trials, with each involving a sample of 16 LLM outputs. Randomly sampling LLM outputs for grading introduces significant variance in alignment across the entire dataset.}
    \label{fig:policy-alignments}
\end{figure}

The findings, shown in \Cref{fig:policy-alignments}, reveal that the random sampling policy exhibits a large variance in alignment. This inconsistency could lead to user frustration, particularly if the effort spent in grading outputs results in assertion sets with unpredictable relevance to their specified criteria. The alternative sampling strategies, which weight the probability for an LLM output to be graded by the selectivity (i.e., pass rate) of assertions that fail it, consistently yielded higher alignment scores across the entire datasets than the random policy. Notably, while the alternating policy didn't consistently outperform, our results suggest that any non-random policy implemented in \evalgen may achieve satisfactory outcomes. 

In this offline study, as shown in \Cref{fig:policy-alignments}, there's no variation in the outcomes of the non-random policies because they are deterministic. However, in real-world use, \evalgen updates its predictions as it receives new information, so there could be some differences in results over time. Initially, when users start grading outputs in \evalgen, they might effectively be grading random outputs for the first one or two outputs, as the $\sigma$ scores update and stabilize.

\section{Task Prompts}\label{appendix:prompts}

We prepared two prompts and corresponding datasets for tasks to present users~(\ref{eval-setup}). Both pipelines were adapted from prior work~\cite{yim2023aci, hou2024bridging} and correspond to medical record processing and a product description writing, respectively. The medical pipeline prompt is as follows:

\begin{promptListing}
You are extracting insights from some medical records. The records contain a medical note and a dialogue between a doctor and a patient. You need to extract values for the following: Chief complaint, History of present illness, Physical examination, Symptoms experienced by the patient, New medications prescribed or changed, including dosages (N/A if not provided), and Follow-up instructions (N/A if not provided). Your answer should not include any personal identifiable information (PII) such as name, age, gender, or ID. Use "the patient" instead of their name, for example. Return your answer as a bullet list, where each bullet is formatted like `chief complaint: xx.` If there is no value for the key, the value should be `N/A`. Keep your response around 150 words (you may have to summarize some extracted values to stay within the word limit).

{transcript}
\end{promptListing}

And the product pipeline prompt is as follows:

\begin{promptListing}
You are an expert copywriter. You need to write an e-commerce product description based on the product details and customer reviews. Your description should be SEO-optimized. It should use an active voice and include the product's features, benefits, unique selling points without overpromising, and a call to action for the buyer. Benefits describe how product features will work for the buyer, addressing exactly how the product will improve their lives. Clearly distinguish between features (e.g., lightweight, USB-chargeable) and benefits (e.g., convenience, nutritious drinks on-the-go). Don't mention weaknesses of the product or use generic or repetitive language. Don't make up review text or quotes. Don't include any links. Don't cite the reviews too heavily. Divide your description into readable chunks divided by relevant subheadings. Keep your description around 200 words, no more than 300, in Markdown format.

{document}
\end{promptListing}




\end{document}